\documentclass[11pt]{article}
\usepackage{latexsym,psfig,rotate,epsf,a4} 

\begin{document}

\begin{center}
{\Large \bf Comparison 
of two models for bridge-assisted charge transfer}

M. Schreiber,
D. Kilin,
 and  
U. Kleinekath\"ofer

{\it Institut f\"ur Physik, Technische Universit\"at, 
D-09107 Chemnitz, Germany}
\end{center}
\date{}

\begin{abstract}
  Based on the reduced density matrix method, we compare two different
  approaches to calculate the dynamics of the electron transfer in
  systems with donor, bridge, and acceptor.  In the first approach a
  vibrational substructure is taken into account for each electronic
  state and the corresponding states are displaced along a common
  reaction coordinate.  In the second approach it is assumed that
  vibrational relaxation is much faster than the electron transfer and
  therefore the states are modeled by electronic levels only. In both
  approaches the system is coupled to a bath of harmonic oscillators
  but the way of relaxation is quite different.  The theory is applied
  to the electron transfer in ${\rm H_2P}-{\rm ZnP}-{\rm Q}$ with
  free-base porphyrin (${\rm H_2P}$) being the donor, zinc porphyrin
  (${\rm ZnP}$) being the bridge and quinone (${\rm Q}$) the acceptor.
  The parameters are chosen as similar as possible for both approaches
  and the quality of the agreement is discussed.
\end{abstract}


\section{Introduction}

Long-range electron transfer (ET) is a very actively studied area in
chemistry, biology, and physics; both in biological and synthetic
systems.  Of special interest are systems with a bridging molecule
between donor and acceptor. For example the primary step of charge
separation in the bacterial photosynthesis takes place in such a
system \cite{bixo91}.  But such systems are also interesting for
synthesizing molecular wires \cite{davi98}.  It is known that the
electronic structure of the bridge component in donor-bridge-acceptor
systems plays a critical role \cite{wasi92,barb96}.  When the bridge
energy is much higher than the donor and acceptor energies, the bridge
population is close to zero for all times and the bridge site just
mediates the coupling between donor and acceptor.  This mechanism is
called superexchange and was originally proposed by Kramers
\cite{kram34} to describe the exchange interaction between two
paramagnetic atoms spatially separated by a nonmagnetic atom. In the
opposite limit when donor and acceptor as well as bridge energies are
closer than $\sim k_{\rm B} T$, the bridge site is actually populated and
the transfer is called sequential.  The interplay between these two
types of transfer has been investigated theoretically in various
publications \cite{bixo91,sumi96,felt95,okad98,schr98}.

In the present work we  compare two different approaches based
on the reduced density matrix formalism. 
In the first model
one pays attention to the fact that experiments in systems
similar to the one discussed here show vibrational coherence
\cite{vos93,stan95}. Therefore 
a vibrational substructure is introduced for each
electronic level  within a multi-level 
Redfield theory \cite{may92,kueh94}.
Below we call this the vibronic model.
In the second approach only
electronic states are taken into account because it is assumed that
the vibrational relaxation is much faster than the ET.
This model is referred to as tight-binding model below.  
In this case
only the relaxation between the electronic states remains.  Such a kind
of relaxation has been phenomenologically introduced for ET by Davis
et al.\ \cite{davi97} and very recently derived in our group
\cite{schr98b,kili99} as a second order perturbation theory in the
system-bath interaction similar to Redfield theory.  
The vibronic and the tight-binding model are described in the
next section and compared in Section 3.

\section{Theory}

For the description of charge transfer and other dynamical processes
in the system we introduce the Hamiltonian
\begin{equation}
  \label{1}
  \hat{H}= \hat{H}_{\rm S}+ \hat{H}_{\rm B}+ \hat{H}_{\rm SB},
\end{equation}
where $\hat{H}_{\rm S}$ denotes the relevant system, $\hat{H}_{\rm B}$
the dissipative bath, and $\hat{H}_{\rm SB}$ the interaction between
the two.  Before discussing the system part of the Hamiltonian in
Sections 2.1 and 2.2, we describe the bath and the procedure how to
obtain the equations of motion for the reduced density matrix, because
this is the same for both models studied below.  The bath is modeled
by a distribution of harmonic oscillators and characterized by its
spectral density $J(\omega)$.  Starting with a density matrix of the
full system, the reduced density matrix of the relevant (sub)system is
obtained by tracing out the bath degrees of freedom \cite{blum96}.
While doing so a second-order perturbation expansion in the
system-bath coupling and the Markov approximation have been applied
\cite{blum96}. 

\subsection{Vibronic model}
The bridge ET system ${\rm H_2P}-{\rm ZnP}-{\rm Q}$ with free-base
porphyrin (${\rm H_2P}$) being the donor, zinc porphyrin (${\rm ZnP}$)
the bridge, and quinone (${\rm Q}$) the acceptor is modeled by
three diabatic electronic potentials,
corresponding to the neutral excited electronic state
$\left|1 \right>=\left|{\rm H_2P}^*-{\rm ZnP}-{\rm Q}\right>$,
and states with charge separation
$\left|2 \right>=\left|{\rm H_2P}^+-{\rm ZnP}^--{\rm Q}\right>$,
$\left|3 \right>=\left|{\rm H_2P}^+-{\rm ZnP}-{\rm Q^-}\right>$
(see Fig.\ 1).
Each of these electronic potentials has a vibrational substructure.
The vibrational frequency is assumed to be 1500 cm$^{-1}$ as a typical
frequency within carbon structures.  The potentials are displaced
along a common reaction coordinate which represents the solvent
polarization \cite{marc56}.  Following the reasoning of Marcus
\cite{marc56} the free energy differences $\Delta G_{mn}$ 
corresponding to the electron transfer from molecular block $n$ to  $m$
($ n=1$, $m=2,3$)
are estimated to be \cite{fuch96d,remp95}
\begin{equation}
  \label{2}
\Delta G_{mn} =E_m^{\rm ox}-E_n^{\rm red}-E^{\rm ex}-
\frac{e^2}{4\pi\epsilon_0\epsilon_{\rm s}}\frac{1}{r_{mn}}+\Delta G_{mn}(\epsilon_{\rm s})
\end{equation}
 with the  term $\Delta G_{mn}(\epsilon_{\rm s})$ correcting for the fact that the redox energies $E^{\rm ox}_m$ and $E^{\rm red}_n$ are measured in a reference solvent with dielectric constant $\epsilon_{\rm s}^{\rm ref}$: 
 \begin{equation}
   \label{2a}
   \Delta G_{mn}(\epsilon_{\rm s})= \frac{e^2}{4 \pi \epsilon_0} 
\left(\frac1{2r_m}+\frac1{2r_n} \right) 
\left( \frac1{\epsilon_{\rm s}^{\rm }} -\frac1{\epsilon_{\rm s}^{\rm ref}} \right).
 \end{equation}
 The excitation energy of the donor ${\rm H_2P} \to {\rm H_2P}^*$ is
 denoted by $E^{\rm ex}$. $r_n$ denotes the radius of
either donor (1), bridge (2), or acceptor (3) 
 and $r_{mn}$ the distance between two of them. They
 are estimated to be $r_1=r_2=5.5$ ${\rm \AA{}}$, $r_3=3.2$ ${\rm \AA{}}$, 
$ r_{12}=12.5$ ${\rm \AA{}}$, and $
 r_{13}=14.4$ ${\rm \AA{}}$ \cite{fuch96d,remp95}.

Also sketched in Fig.\ 1 are the reorganization energies 
$\lambda_{mn}=\lambda_{mn}^{\rm i}+\lambda_{mn}^{\rm s}$.
These consist of an internal reorganization energy $\lambda_{mn}^{\rm i}$, 
which is
estimated to be 0.3 eV \cite{remp95}, and a solvent reorganization energy
\cite{marc56}
\begin{equation}
  \label{3}
  \lambda_{mn}^{\rm s}=\frac{e^2}{4 \pi \epsilon_0}
\left(\frac1{2r_m}+\frac1{2r_n}-\frac1{r_{mn}}\right) 
\left(\frac1{\epsilon_{\infty}}-\frac1{\epsilon_{\rm s}}\right)~.
\end{equation}
Further parameters are the electronic couplings between the potentials.
First it should be underlined that $V_{13}=0$ because of the
spatial separation of ${\rm H_2P}$ and ${\rm Q}$.
So there is no direct transfer between donor and acceptor.
 The other couplings are
$V_{12}=65$ ${\rm meV}$ and $V_{23}=2.2$ ${\rm meV}$ \cite{remp95}.
The damping is described by  the
spectral density $J(\omega)$ of the bath. This is only needed at the 
frequency of the vibrational transition and is determined
$J(\omega_{\rm vib})/\omega_{\rm vib}=0.372$ 
by fitting the ET rate for the solvent methyltetrahydrofuran (MTHF).
In the vibronic model the  spectral density
is taken as a constant with respect to $\epsilon_{\rm s}$.

Next the calculation of the dynamics is sketched.
Starting from the Liouville equation, performing 
the abovementioned approximations
the equation of motion for the
reduced density matrix $\rho_{\mu{}\nu}$ can be obtained \cite{may92,kueh94}
\begin{equation}
  \label{10}
  \frac{\partial}{\partial t} \rho_{\mu{}\nu}=\frac{i}{\hbar}
(E_\mu{}-E_{\nu}) \rho_{\mu{}\nu} - i \sum_\kappa \{ v_{\nu \kappa} 
\rho_{\mu \kappa} -v_{\kappa \mu} \rho_{\kappa \nu} \} +R_{\mu{}\nu}~.
\end{equation}
The index $\mu{}$ combines the electronic quantum number $m$ and the
vibrational quantum number $M$ of the diabatic levels $E_\mu{}$.
$v_{\mu{}\nu}=V_{mn} F_{\rm FC}(m,M,n,N)$ comprises Franck-Condon
factors $F_{\rm FC}$ and the electronic matrix elements $V_{mn}$.  The third
term describes the interaction between the relevant system and the
heat bath.
Equation~(\ref{10}) is solved numerically with the initial condition that only the
donor state is occupied in the beginning.
The population of the acceptor state
\begin{equation}
P_3(t)=\sum \limits_M \rho_{3M3M}(t)
\end{equation}
and the ET rate
\begin{equation}
k_{\rm ET}=\frac{P_3(\infty)}{\int \limits_0^\infty dt(1-P_3(t))}
\end{equation}
are calculated by tracing out the vibrational modes.

\subsection{Tight-binding model}

The reasoning for the following system Hamiltonian is the assumption that
the vibrational excitations are relaxed on a much shorter time scale than
the ET time scale. Therefore only electronic states without any vibrational
substructure are taken into account (see Fig.\ 2). As a consequence 
the relaxation during the ET process has to be described in a different
manner than in the previous subsection. If now relaxation takes place, it
takes place between the electronic states and not between vibrational
states within one electronic state potential surface.  A similar model has
been introduced phenomenologically by Davis et al.\ \cite{davi97} who
solved it in the steady state limit.

The energies of the electronic states $E_m$ are chosen to be
the ground states of the harmonic potentials given in the previous section.
So they vary with the dielectric constant. 
The  electronic coupling is fixed in two different ways. In the naive
way they are chosen to be the same as in the vibronic model.
But because in the tight-binding model there is no reaction coordinate,
in a second version we scale the electronic couplings with
the Franck-Condon overlap elements between the vibrational ground
states of each pair of electronic surfaces
\begin{equation}
  \label{5}
v_{mn}=V_{mn}F_{\rm FC}(m,0,n,0)
=V_{mn}\exp{\frac{-|\lambda_{mn}|}{2 \hbar \omega_{\rm vib}}}~.
  \end{equation}
  In the vibronic model not only the free energy differences $\Delta
  G$ but also the reorganization energies $\lambda$ scale with the
  dielectric constant $\epsilon_{\rm s}$.  Due to this scaling of
  $\lambda$ the system-bath interaction is scaled with the dielectric
  constant $\epsilon_{\rm s}$. In the high temperature limit the
  reorganization energy is given by \cite{weis99}
\begin{equation}
  \label{4}
  \lambda = \hbar \int_0^{\infty} d\omega\frac{J(\omega)}{\omega}~.
\end{equation}
This relation is taken as motivation to
 scale the tight-binding spectral density
with  $\epsilon_{\rm s}$ like the reorganization energies $\lambda$
in the vibronic model. 
In the present calculations
$\Gamma_{21}=\Gamma_{23}=\Gamma$ is assumed. 
The absolute value of the damping rate $\Gamma$ between the 
electronic states  (see Fig.~2) is then determined by fitting the ET
rate for the solvent MTHF to be $\Gamma=2.8\times{}10^{11}$ s$^{-1}$.
  
The advantage of the tight-binding model is the possibility to
determine the transfer rate $k_{\rm ET}$ and the final population of
the acceptor state either numerically or analytically.  We employ the
rotating wave approximation because we are only interested in the
reaction rates here.  For the analytic calculation three extra
assumptions have to be made: small bridge population, the kinetic
limit $t\gg{}\Gamma^{-1}$, and the absence of initial coherence in the
system.  But for all situations described in this paper the
differences between analytic and numerical results without the extra
assumptions are negligible.  The analytic expressions are
\begin{equation} 
\label{rate}
k_{\rm ET}=g_{23}+\frac{g_{23}(g_{12}-g_{32})}{g_{21}+g_{23}} 
\end{equation}
and 
\begin{equation}
\label{population}
P_3(\infty)=\frac{g_{12}g_{23}}{g_{21}+g_{23}}(k_{\rm ET})^{-1}, 
\end{equation} 
which contain both, dissipative and coherent contributions
\begin{equation} 
\label{dissipation}
g_{mn}= 
d_{mn}
+
\frac{v_{mn}^2 \sum\limits_k (d_{mk}+d_{kn})}
{\hbar^2\left\{2\omega^2_{mn}+\frac12\left[\sum\limits_k (d_{mk}+d_{kn}) \right]^2\right\}}. 
\end{equation}
Herein the $d_{mn}$ are just abbreviations for $\Gamma_{mn}
|n(\omega_{mn})|$ and $n(\omega_{mn})$ denotes the Bose distribution
at frequency $\omega_{mn}=(E_m-E_n)/\hbar$.  For details and
comparison with the Grover-Silbey theory \cite{silb71} as well as the
Haken-Strobl-Reineker theory \cite{rein82} we refer the reader to Ref.\ 
\cite{kili99}.

\section{Comparison}
In Fig.\ 3 it is shown how the minima of the potential curves change
with varying the solvent due to the changes in Eqs.\ (\ref{2}) to
(\ref{3}). The solvents are listed
in Table 1 together with their parameters and the results for the ET
rates in both models.
For larger $\epsilon_{\rm s}$ the coordinates of the potential
minima of bridge and acceptor increase while their energies decrease
with respect to the energy of the donor.  The energy
difference between donor and bridge decreases with increasing
$\epsilon_{\rm s}$.  This makes a charge transfer more probable.
For small $\epsilon_{\rm s}$ 
the acceptor state is higher in energy than the donor state;
nevertheless there is a small ET rate due to coherent mixing.

For fixed $\epsilon_{\infty}$ the ET rate is plotted as a function of
the dielectric constant $\epsilon_{\rm s}$ in Fig.\ 4. The ET rate in
the vibronic model increases strongly for small values of
$\epsilon_{\rm s}$ while the increase is very small for $\epsilon_{\rm
  s}$ in the range between 5 and 8. The increase for small values of
$\epsilon_{\rm s}$ is due to the fact that with increasing
$\epsilon_{\rm s}$ the minimum of the acceptor potential moves from a
position higher than the minimum of the donor level to a position
lower than the donor level. So the transfer becomes energetically
favorable.  This can also be seen when looking at the results for the
tight-binding model without scaling the electronic coupling with the
Franck-Condon factor. In this case the ET rate increases almost
linearly with increasing $\epsilon_{\rm s}$. The effect missing in
this model is the overlap between the vibrational states. If one
corrects the electronic coupling in the tight-binding model by the
Franck-Condon factor of the vibrational ground states as described in
Eq.\ (\ref{5}), good agreement is observed between the vibronic and
the tight-binding model.

The ET rate for the vibronic model shows some oscillations as a function of
$\epsilon_{\rm s}$. This is due to the small density of vibrational levels
in this model with one reaction coordinate. All three electronic potential
curves are harmonic and have the same frequency.  So there are small maxima
in the rate when two vibrational levels are in resonance and minima when
they are far off resonance. Models with more reaction coordinates do not
have this problem nor does the simple tight-binding model. If these
artificial oscillations would be absent, the agreement between the results
for the tight-binding and the vibronic model would be even better, because
the rate for the vibronic model happens to have a maximum just at the
reference point $\epsilon_{\rm s}=6.24$ which we have chosen to fix the spectral density, i.\ e.\ 
for MTHF.

The comparison of the two models has been made assuming that the scaling of
energies as a function of the dielectric function is correct in the Marcus
theory. There have been a lot of changes to Marcus theory proposed in the
last years.  Marcus theory assumes excess charges within cavities
surrounded by a polarizable medium and there one only takes the leading
order into account.  Higher order terms are included in the so called
reaction field theory (see for example \cite{kare97}). But to compare
different solvation models is out of the range of the present
investigation.  Some more details on this issue for the tight-binding model
are given in Ref.\ \cite{kili99}. Here we just want to note in passing that the
effect of scaling the system-bath interaction with $\epsilon_{\rm s}$, as
assumed in the present work for the tight-binding model, has no big effect
on the ET rates.

As conclusion we mention that one gets good agreement for the ET rates of
the models with and without vibrational substructure, i.\ e.\ the vibronic
and the tight-binding model, if one scales the electronic coupling with the
Franck-Condon overlap matrix elements between the vibrational ground
states.  The advantage of the model with electronic relaxation only is the
possibility to derive analytic expressions for the ET rate and the final
population of the acceptor state.  But of course for a more realistic
description of the ET transfer process in such complicated systems as
discussed here, more than one reaction coordinate should be taken into
account.  Work in this direction is in progress.

\section{Acknowlegements}
  We thank I. Kondov for the help with some programming as well as
  U.~\mbox{Rempel} and E.~Zenkevich for stimulating discussions.
  Financial support of the DFG is gratefully acknowledged.

\newpage

\tabcolsep=0.05cm
\begin{table}[htp]
  \begin{center}
    \leavevmode
\begin{tabular}{l|c|c|c|c|c|c|c|c|c}
\hline
solvent& $\epsilon_{\rm s}$&$\epsilon_\infty$&$\Delta G_{21}$&$\Delta G_{31}$&$\lambda_{21}^{\rm s}$&$\lambda_{31}^{\rm s}$&$\Gamma$ &$k_{\rm ET}^{\rm el}$ &$k_{\rm ET}^{\rm vib}$  \\
& && [{\rm eV}]& [{\rm eV}]&[{\rm eV}]&[{\rm eV}]&[$10^{11}$ s$^{-1}$]& [$10^8$ s$^{-1}$]&[$10^8$ s$^{-1}$] \\
\hline
1. cyclohexane \protect \cite{remp95}             &   2.02 & 2.00 & 0.976 &    0.393 &  0.007 & 0.012 & 0.042 &  0.181 & 0.7   \\
2. toluene \protect \cite{tenn99}                 &   2.38 & 2.24 & 0.867 &    0.202 &  0.039 & 0.069 & 0.227 &  1.04  & 0.8   \\
3. anisole  \protect \cite{schm89}                &   4.33 & 2.29 & 0.590 &   -0.281 &  0.300 & 0.524 & 1.751 &  4.24  & 2.30  \\
4. dibromoethane    \protect \cite{schm89}        &   4.78 & 2.37 & 0.558 &   -0.336 &  0.312 & 0.544 & 1.817 &  4.63  & 2.45  \\
5. chlorobenzene     \protect \cite{tenn99}       &   5.29 & 1.93 & 0.529 &   -0.388 &  0.481 & 0.839 & 2.804 &  3.21  & 3.63  \\
6. MTHF    \protect \cite{remp95}                 &   6.24 & 2.00 & 0.486 &   -0.462 &  0.497 & 0.868 & 2.900 &  3.59  & 3.58  \\
7. methyl acetate    \protect \cite{tenn99}       &   6.68 & 1.85 & 0.471 &   -0.489 &  0.571 & 0.996 & 3.328 &  2.96  & 4.15  \\
8. trichloroethane      \protect \cite{schm89}    &   7.25 & 2.06 & 0.454 &   -0.512 &  0.508 & 0.887 & 2.960 &  3.98  & 3.50   \\
9. dichloromethane \protect \cite{remp95}         &   9.08 & 2.03 & 0.413 &   -0.590 &  0.559 & 0.977 & 3.264 &  4.00  & 3.80  \\
\hline
\end{tabular}
        \caption{Parameters and obtained transfer rates 
 for different solvents. The references behind
the names of the solvents cite the sources of $\epsilon_{\rm s}$
and $\epsilon_\infty$.
MTHF stands for methyltetrahydrofuran. 
$\Gamma$ denotes the damping rate in the tight-binding model.
The ET rate for the solvent MTHF
has been used to fix the damping parameter of the models.
The reaction rates $k_{\rm ET}^{\rm el}$ were obtained using
\protect Eq.~(\ref{rate})
within the tight-binding model and
the reaction rates $k_{\rm ET}^{\rm vib}$  within the vibronic model.
}
    \label{tab1}
  \end{center}
\end{table} 

\newpage

\begin{figure}[htp]
  \begin{center}
\parbox{6.0cm}{\rule{-3cm}{.1cm}\epsfxsize=11.0cm\epsfbox{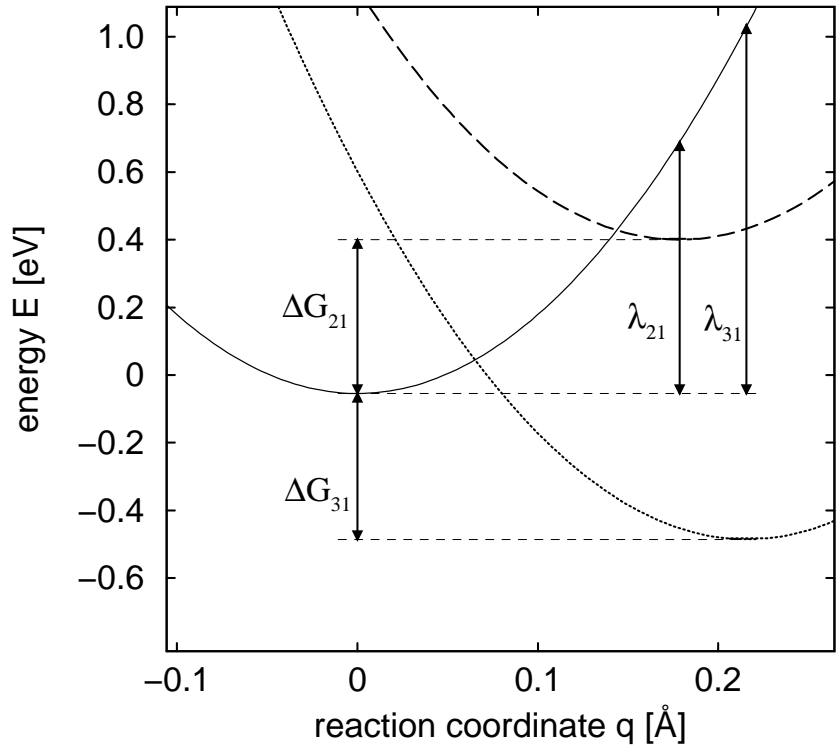}}
    \leavevmode
    \caption{Electronic potentials and parameters of the vibronic model.
      The donor surface $\left|{\rm H_2P}^*-{\rm ZnP}-{\rm Q}\right>$
      is given by the solid line, the bridge $\left|{\rm H_2P}^+-{\rm
          ZnP}^--{\rm Q}\right>$ by the dashed line, and the acceptor
      $\left|{\rm H_2P}^+-{\rm ZnP}-{\rm Q^-}\right>$ by the dotted
      line. }
    \label{fig1}
  \end{center}
\end{figure}
 
\begin{figure}[htp]
  \begin{center}
\parbox{6.0cm}{\rule{-3cm}{.1cm}\epsfxsize=11.0cm\epsfbox{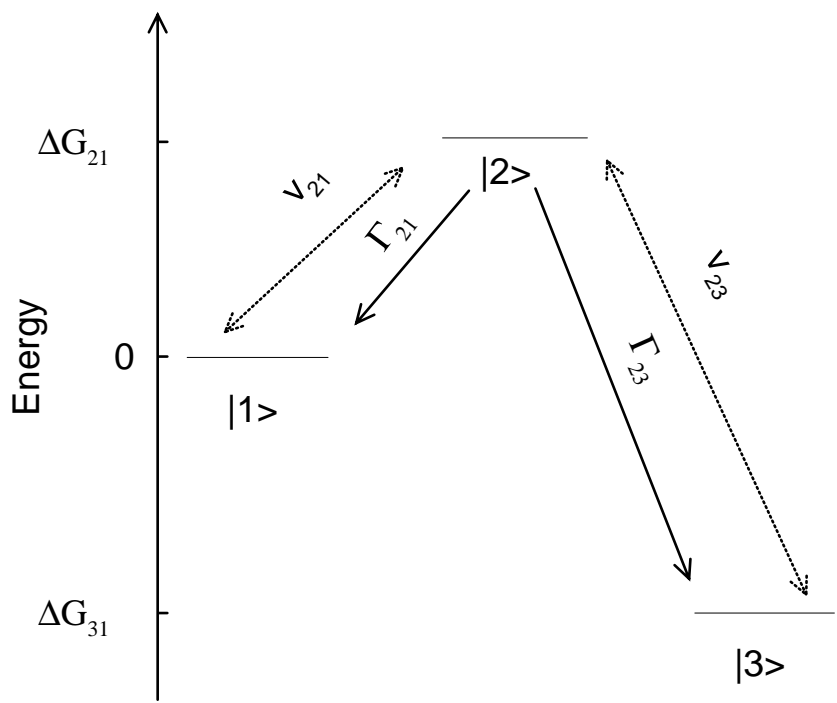}}
    \leavevmode
    \caption{Schematic presentation of the tight-binding model.}
    \label{fig2}
  \end{center}
\end{figure}

\begin{figure}[h]
  \begin{center}
\parbox{6.0cm}{\rule{-3cm}{.1cm}\epsfxsize=11.0cm\epsfbox{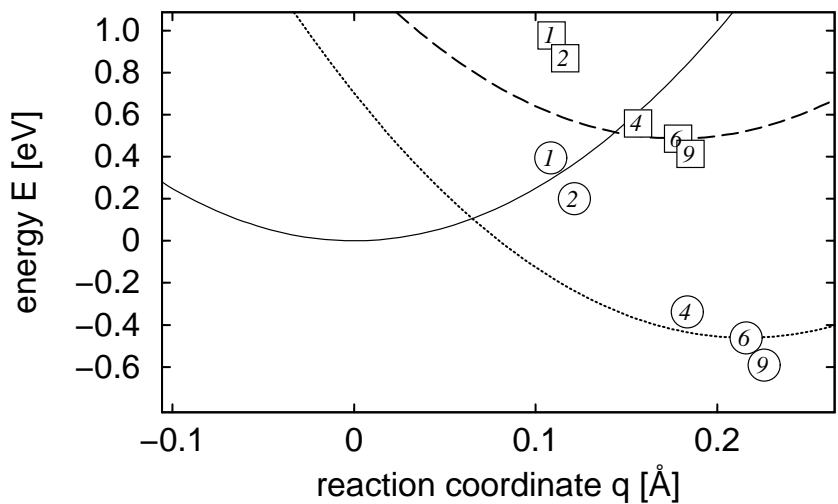}}
    \leavevmode
    \caption{Variation of the potential minima for different solvents.
Squares denote the bridge minima, circles the acceptor minima.
The numbers correspond to the ordinal numbers in Table 1. The potentials
are shown for solvent 6  (MTHF). }
    \label{fig3}
  \end{center}
\end{figure}

\begin{figure}[htp]
  \begin{center}
\parbox{6.0cm}{\rule{-3cm}{.1cm}\epsfxsize=11.0cm\epsfbox{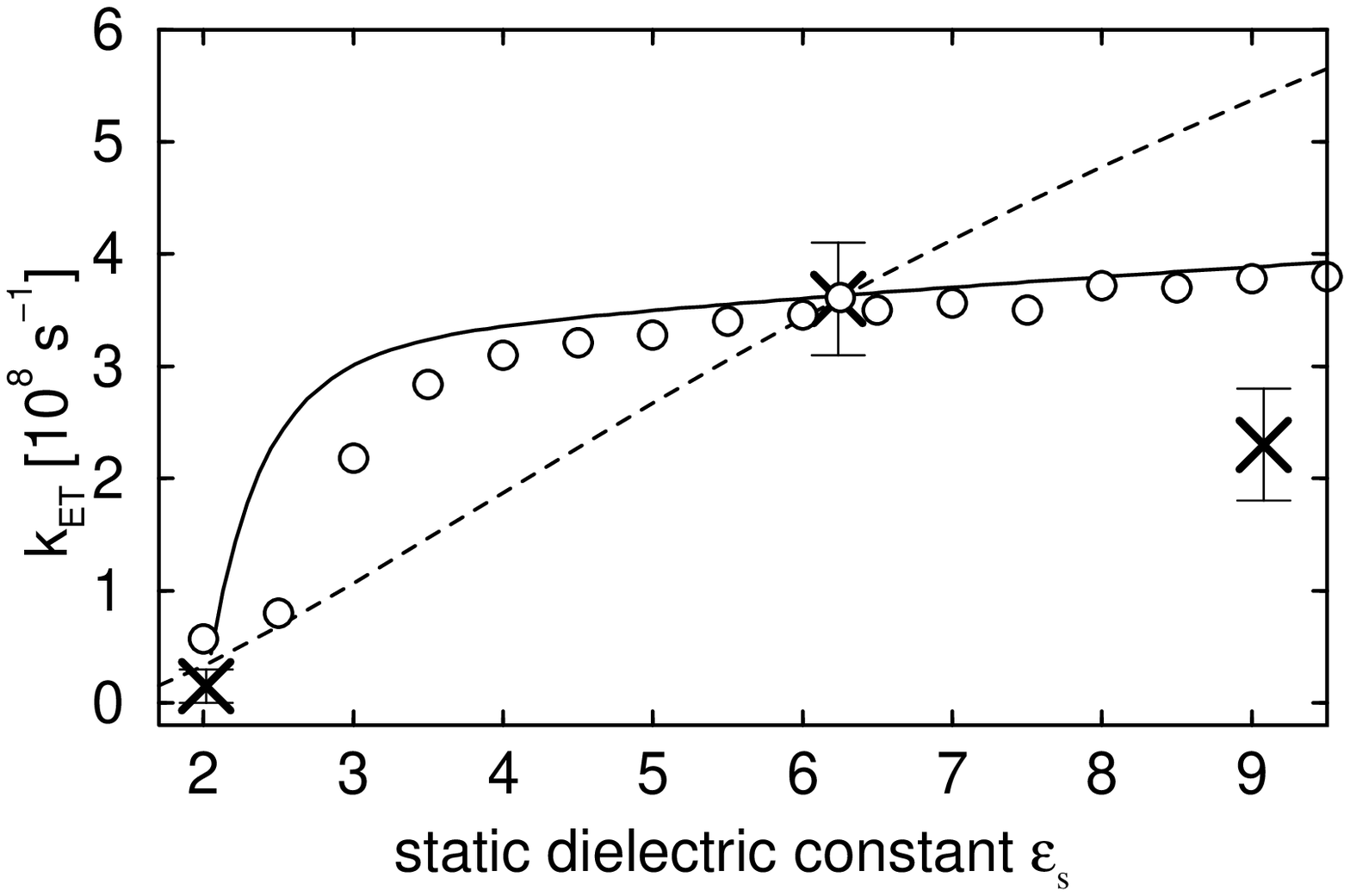}}
    \leavevmode
    \caption{Transfer rate as a function of the dielectric constant 
   $\epsilon_{\rm s}$ for both models together with experimental 
results \protect \cite{remp95}. The rates for the vibronic model
are given by the circles.
The dashed line shows the rate for the tight-binding model  with
electronic couplings $V_{mn}$ as in the vibronic model. The
solid line represents the rate for the tight-binding model
with $v_{mn}$ scaled as given in Eq.\ \protect (\ref{5}).}
    \label{fig4}
  \end{center}
\end{figure}


\begin{thebibliography}{10}
\bibitem{bixo91} M. Bixon, J. Jortner and M. E. Michel-Beyerle,
  Biochim.\ Biophys.\ Acta 1056 (1991) 301; Chem.\ Phys.\ 197 (1995)
  389.  
\bibitem{davi98} W. B. Davis, W. A. Svec, M. A. Ratner, and M.
  R.  Wasielewski, Nature 396 (1998) 60.\ 
\bibitem{wasi92} M.~R.
  Wasielewski, Chem. Rev.  92 (1992) 345.  
\bibitem{barb96} P.~F.
  Barbara, T.~J. Meyer, and M.~A.~Ratner, J. Phys. Chem.  100 (1996)
  13148.  
\bibitem{kram34} H. A. Kramers, Physica 1 (1934) 182.
\bibitem{sumi96} H.~Sumi and T.~Kakitani, Chem.\ Phys.\ Lett.\ 252
  (1996) 85; H.~Sumi, J. Electroan.\ Chem.\ 438 (1997) 11.
\bibitem{felt95} A.~K.~Felts, W.~T.~Pollard, and R.~A.~Friesner, J.
  Phys.\ Chem.\ 99 (1995) 2929.  
\bibitem{okad98} A. Okada, V.
  Chernyak, and S. Mukamel, J. Phys.\ Chem.\ A 102 (1998) 1241.
\bibitem{schr98} M. Schreiber, C. Fuchs, and R.~Scholz, J. Lumin.\ 
  76\&77 (1998) 482.  
\bibitem{vos93} M.~H.~Vos, F.~Rappaport,
  J.-C.~Lambry, J.~Breton, and J.-L.~Martin, Nature 363 (1993) 320.
\bibitem{stan95} R.~J.~Stanley and S.~G.~Boxer, J. Phys.\ Chem.\ 99
  (1995) 859.  
\bibitem{may92} V.~May and M.~Schreiber, Phys.\ Rev.\ A
  45 (1992) 2868.  \bibitem{kueh94} O.~K\"uhn, V.~May, and
  M.~Schreiber, J.\ Chem.\ Phys.\ 101 (1994) 10404.  
\bibitem{davi97}
  W.~Davis, M.~Wasilewski, ~M.  Ratner, V.~Mujica, and A. Nitzan, J.
  Phys. Chem.  101 (1997) 6158.  
\bibitem{schr98b} M. Schreiber, D.
  Kilin, and U. Kleinekath\"ofer, in: R.~T.~Williams and W.~M.~Yen
  (Eds.), Excitonic Processes in Condensed Matter, PV 98-25, p.~99,
  The Electrochemical Society Proceedings Series, Pennington, NJ,
  1998.  
\bibitem{kili99} D.  Kilin, U. Kleinekath\"ofer, and M.
  Schreiber (in preparation).  
\bibitem{blum96} K. Blum, Density
  Matrix Theory and Applications, Plenum Press, New York, 1996, 2nd
  ed.\ 
\bibitem{marc56} R.~A.~Marcus, J.\ Chem.\ Phys.\, 24 (1956)
  966; R.~A.~Marcus und N.~Sutin, Biochim.\ Biophys.\ Acta 811 (1985)
  265.  
\bibitem{fuch96d} C. Fuchs, Ph.D. thesis, Technische
  Universit\"at Chemnitz, 1997,
  http://archiv.tu-chemnitz.de/pub/1997/0009 
\bibitem{remp95}
  U.~Rempel, B.~von~Maltzan, and C.~von~Borczyskowski, Chem. Phys.
  Lett. 245 (1995) 253.  
\bibitem{weis99} U.~Weiss, Quantum
  Dissipative Systems, World Scientific, Singapore, 1999.
\bibitem{silb71} M.~Grover and R.~Silbey, J.\ Chem.\ Phys.\ 54 (1971)
  4843.  
\bibitem{rein82} P. Reineker, in: G. H\"ohler (Ed.), Exciton Dynamics
  in Molecular Crystals and Aggregates, Springer, Berlin, 1982.
\bibitem{kare97} M.~Karelson, G.~H.~F.~Diercksen, in: S.~Wilson and
  G.~H.~F.~Diercksen, Problem Solving in Computational Molecular
  Science: Molecules in Different Environments, Kluwer, Dordrecht,
  1997.  
\bibitem{tenn99} \mbox{Charles Tennant \& Company (London)
    Ltd}, http://www.ctennant.co.uk/tenn04.htm 
\bibitem{schm89}
  J.~A.~Schmidt, J.-Y.~Liu, J.~R.~Bolton, M.~D.~Archer, and
  V.~P.~Y.~Gadzepko, J. Chem. Soc. Faraday Trans. 85 (1989) 1027.
\end{thebibliography}
\end{document}